\documentclass[]{article}
\usepackage{amsmath,amsthm}
\usepackage{eufrak}
\newcommand{\ar}{\renewcommand{\arraystretch}{1}} 
\DeclareMathAlphabet{\bb}{U}{msb}{m}{n}

 \DeclareMathOperator{\spin}{{\bf
Spin}} 

 \DeclareMathOperator{\Ad}{Ad}

\DeclareMathOperator{\SO}{SO} 
\DeclareMathOperator{\SU}{SU} \DeclareMathOperator{\Sp}{Sp}

\newcommand{\cH}{\mathcal{H}}

\newcommand{\cL}{\mathcal{L}}

\newcommand{\cM}{{\cal M}}

\newcommand{\bi}{{\bf i}}

\newcommand{\fM}{\mathfrak{M}}

\newcommand{\fP}{\mathfrak{P}}

\newcommand{\fS}{\mathfrak{S}}
\newcommand{\fZ}{\mathfrak{Z}}

\newcommand{\fq}{\mathfrak{q}}

\newcommand{\cl}{C\kern -0.2em \ell}

\newcommand{\hypergeom}[5]{\mbox{$
_#1 F_#2\left. \!\! \left( \!\!\!\!\ar
\begin{array}{c}
\multicolumn{1}{c}{\begin{array}{c} #3
\end{array}}\\[1mm]
\multicolumn{1}{c}{\begin{array}{c} #4
\end{array}}\end{array}
\!\!\!\! \right|\displaystyle{#5}\right) $} }

\begin{document}
\title{Group averaging for de Sitter free fields in terms of
hyperspherical functions}
\author{V.~V. Varlamov\\
{\small\it Siberian State Industrial University, Novokuznetsk
654007, Russia}}
\date{}
\maketitle
\begin{abstract}
We study the convergence of inner products of free fields over the
homogeneous spaces of the de Sitter group and show that the
convergence of inner products in the of $N$-particle states is
defined by asymptotic behavior of the hypergeometric functions. We
calculate the inner product for two-particle states on the
four-dimensional hyperboloid in detail.
\end{abstract}
\vspace{0.5cm} \textbf{Keywords}: de Sitter group, group averaging,
inner product, hyperspherical function, homogeneous space.

\section{Introduction}
As is known, the procedure introduced by Dirac \cite{Dir} for
quantizing constrained systems is currently intensively studied in
theoretical physics. In the Dirac approach, constraints are
considered operators acting on the vectors of some Hilbert space and
also conditions selecting ``physical'' states, from which a physical
Hilbert space is then formed. But this quantization procedure
contains a number of unsolved problems related to the structure of
the Hilbert space of physical states and to the definition of inner
products for these states. Some progress in this area has been
achieved in such approaches as the BRST-method \cite{HT92}, the
method of geometric quantization \cite{Woo80}, coherent state
quantization \cite{Klau98}, $C^\ast$-algebra methods \cite{GH85},
algebraic quantization \cite{Ash91}, and refined algebraic
quantization (RAQ) \cite{ALMMT95,Mar95}, related with the Rieffel
induction \cite{Lan95,Rie74} and other works (see
e.g.,\cite{Mar95b}). In the framework of RAQ, the inner product of
states is defined using the technique of group averaging. Group
averaging uses the integral
\[
\int_G\langle\phi_1|U(g)|\phi_2\rangle dg
\]
over the gauge group $G$, where $dg$ is a so-called symmetric Haar
measure on $G$, $U(g)$ is a representation of $G$, and $\phi_1$ and
$\phi_2$ are state vectors from an auxiliary Hilbert space
$\cH_{aux}$. Convergent group averaging gives an algorithm for
construction of a complete set of observables of a quantum system
\cite{Hig91,GM99,MM09,Shv02}.

Here, we study inner products of free fields of any spin over
homogeneous spaces of the de Sitter group $\SO_0(1,4)$. The key
point in studying the convergence of group averaging is the method
for defining matrix elements of representations $U(g)$ of
$\SO_0(1,4)$ using an additional theorem for generalized spherical
functions previously developed \cite{Var07,Var04e,Var06,Var05b}. The
main advantage of this way of defining the matrix elements is their
explicit factorization (according to the Cartan decomposition) with
respect to the subgroups of the original group. Hence, for the group
$\SO_0(1,4)$, the matrix elements can be factorized with respect to
both $\SO(4)$ (a maximal compact subgroup) and $\SO_0(1,3)$ (Lorentz
group). This factorization allows segregating the variables in the
integral that defines the group averaging for the inner product,
i.e., calculating the integrals over compact and noncompact
subgroups separately. As an example, we calculate the inner product
in the two-particle case on the four-dimensional hyperboloid in
detail. We show that the convergence of inner products is defined by
the asymptotic behavior of the hypergeometric functions.

\section{Group averaging and inner products over the de Sitter group}
The main idea in the Dirac approach is to impose the additional
conditions
\[
\hat{\Lambda}^+_a\mid\boldsymbol{\Psi}\rangle=0,\quad
a=\overline{1,\,M},
\]
on the wave function $\mid\boldsymbol{\Psi}\rangle$, where
$\hat{\Lambda}^+_a$ are quantum analogues of the constraints. The
commutation relations
\[
\left[\hat{\Lambda}^+_a,\hat{\Lambda}^+_b\right]=\bi(U^c_{ab})
\hat{\Lambda}^+_c
\]
then hold, where operators $U^c_{ab}$ usually called structure
functions. The most difficult problem of the Dirac approach is to
construct the inner product because $\boldsymbol{\Psi}(q)$ are
probability distributions and not square-integrable functions.

An alternative method in quantum theory is the RAQ. The essence of
this method is the introduction of arbitrary functions
$\boldsymbol{\Phi}(q)\in\cH_{aux}$ called auxiliary state vectors,
whose inner product is given by the formula $
(\boldsymbol{\Phi},\eta\boldsymbol{\Phi}), $ where the map $\eta$ is
such that $
\mid\boldsymbol{\Psi}\rangle=\eta\mid\boldsymbol{\Phi}\rangle. $ The
Dirac states $\mid\boldsymbol{\Psi}\rangle$ form the physical
Hilbert space $\cH_{phys}$. In the case of a general closed algebra,
the inner product is expressed via the integral over the gauge group
$G$ using the \emph{group averaging formula}
\[
\int
d_Rg(\det\Ad\{g\})^{-1/2}\langle\boldsymbol{\Phi}\mid\check{T}(g)
\mid\boldsymbol{\Phi}\rangle,
\]
where $d_Rg$ is a right-invariant Haar measure on the group $G$ and
$\check{T}(g)$ is a representation of $G$. In the case of
$\SO_0(1,4)$, we have the inner product
\begin{equation}\label{stat}
\langle\boldsymbol{\Psi}_1\mid\boldsymbol{\Psi}_2\rangle=\int_{g\in
G}dg\langle\boldsymbol{\Phi}_1\mid
U(g)\mid\boldsymbol{\Phi}_2\rangle.
\end{equation}
The convergence of this integral is completely defined by the matrix
elements of irreducible unitary representations of $\SO_0(1,4)$.

It was shown it \cite{Var07} that the matrix elements and spherical
functions of irreducible representations of the de Sitter group
$\SO_0(1,4)$ form a universal covering $\spin_+(1,4)\simeq\Sp(1,1)$
of $\SO_0(1,4)$. Spherical functions on $\SO_0(1,4)$ are understood
as functions of class-1 representations realized on homogeneous
spaces of $\SO_0(1,4)$. A list of homogeneous spaces of $\SO_0(1,4)$
including symmetric Riemannian and non-Riemannian spaces was given
in \cite{Var07}. Matrix elements realized on the $\SO_0(1,4)$ group
manifold $\fS_{10}$  have the form
\[
\fM^l_{mn}(\fq)=e^{-\bi m\varphi^q}\fZ^l_{mn}(\cos\theta^q)e^{-\bi
n\psi^q},
\]
where $l=0,1/2,1,\ldots$, and $-l\leq m,n\geq l$. The hyperspherical
function $\fZ^l_{mn}(\cos\theta^q)$ is expressed as a series in
products of three hypergeometric functions
\begin{multline}
\fZ^l_{mn}(\cos\theta^q)=\sqrt{\frac{\Gamma(l+m+1)\Gamma(l-n+1)}
{\Gamma(l-m+1)\Gamma(l+n+1)}}
\cos^{2l}\frac{\theta}{2}\cos^{2l}\frac{\phi}{2}\cosh^{2l}\frac{\tau}{2}
\times\\
\sum^l_{k=-l}\sum^l_{t=-l}\bi^{m-k}\tan^{m-t}\frac{\theta}{2}\tan^{t-k}
\frac{\phi}{2}
\tanh^{k-n}\frac{\tau}{2}
\hypergeom{2}{1}{m-l,-t-l}{m-t+1}{-\tan^2\frac{\theta}{2}}\times\\
\hypergeom{2}{1}{t-l,-k-l}{t-k+1}{-\tan^2\frac{\phi}{2}}
\hypergeom{2}{1}{k-l,-n-l}{k-n+1}{\tanh^2\frac{\tau}{2}} \nonumber
\end{multline}
for $m\geq t,\;t\geq k,\;k\geq n$. There also exist seven
expressions of the hypergeometric type for the functions
$\fZ^l_{mn}(\cos\theta^q)$ with the index values $m\geq t,\;k\geq
t$, and $k\geq n$; $t\geq m,\;k\geq t$, and $n\geq k$; $t\geq
m,\;t\geq k$, and $n\geq k$; $t\geq m,\;k\geq t$, and $k\geq n$;
$t\geq m,\;t\geq k$, and $k\geq n$; $m\geq t,\;t\geq k$, and $n\geq
k$; and $m\geq t,\;k\geq t$, and $n\geq k$.

\section{Convergence of inner products}
Returning to the group-averaging formula, we see that the
convergence of the inner product
\[
\langle\boldsymbol{\Psi}_1\mid\boldsymbol{\Psi}_2\rangle:=\int_{g\in
G}dg\langle\phi_1\mid U(g)\mid\phi_2\rangle
\]
depends on the matrix elements of irreducible representations $U(g)$
of the de Sitter group $\SO_0(1,4)$. It was previously shown
\cite{Var07} that such matrix elements are defined via the
hyperspherical function of the form
\[
\mathfrak{M}^l_{mn}(\mathfrak{q})=e^{-\boldsymbol{i}(m\varphi^q+n\psi^q)}
\sum^l_{k=-l}\sum^l_{t=-l}
P^l_{mk}(\cos\phi)P^l_{kt}(\cos\theta)\mathfrak{P}^l_{tn}(\cosh\tau),
\]
where $P^l_{mk}(\cos\phi)$, $P^l_{kt}(\cos\theta)$ are spherical
functions on the subgroup $\SU(2)$ and
$\mathfrak{P}^l_{tn}(\cosh\tau)$ is a spherical function on the
subgroup $\SU(1,1)$. This expression follows directly from the
Cartan decomposition $U(g)=A^qK^qA^q$, where
$K^q=\SU(2)\otimes\SU(2)$ and $A^q$ are maximal compact and
commutative subgroups of $\Sp(1,1)$. With this expression taken into
account, group averaging inner product (\ref{stat}) on the group
manifold $\fS_{10}$ becomes
\[
\langle\boldsymbol{\Psi}_1\mid\boldsymbol{\Psi}_2\rangle:=
\int\limits^\infty_0d\tau
d\epsilon d\varepsilon
d\omega\sin\theta^qe^{-m\epsilon-n(\varepsilon+\omega)}\langle\psi_1\mid
P_0K^qP_0\mid\psi_2\rangle,
\]
where $P_0$ are projectors on $\SU(2)$-invariant states. Because
$\SU(2)$ is a compact group, the operators $P_0$ do not affect the
convergence properties of the group-averaging inner product. For
$N$-particle states in which each particle occupies a definite mode,
we have
\[
\langle\boldsymbol{\Psi}_1\mid\boldsymbol{\Psi}_2\rangle:=
\int\limits^\infty_0d\tau
d\epsilon d\varepsilon
d\omega\sin\theta^qe^{-m\epsilon-n(\varepsilon+\omega)}
\mathfrak{P}^l_{m_1,n_1}(\cosh\tau)\cdots
\mathfrak{P}^l_{m_N,n_N}(\cosh\tau).
\]
The convergence of this integral follows from asymptotic behavior of
the hypergeometric type functions
\begin{multline}
\fP^l_{mn}(\cosh\tau)=\frac{1}{\Gamma(m-n+1)}
\sqrt{\frac{\Gamma(l-n+1)\Gamma(l+m+1)}{\Gamma(l-m+1)\Gamma(l+n+1)}}\times\\
\cosh^{m+n}\frac{\tau}{2}\sinh^{m-n}\frac{\tau}{2}
\hypergeom{2}{1}{l+m+1,m-l}{m-n+1}{-\sinh^2\frac{\tau}{2}}.
\label{Jacobi}
\end{multline}

As an example, we consider the inner product of two-particle states
on the homogeneous space $\cM_4=\SO_0(1,4)/\SO(4)$. This space is
homeomorphic to a two-sheeted four-dimensional hyperboloid $H^4$. We
note that \emph{a four-dimensional Lobatchevski space} $\cL^4$, also
called \emph{a de Sitter space}, is realized on the hyperboloid
$H^4$. Spherical functions defined on the homogeneous space
$\cM_4=H^4_+\sim\SO_0(1,4)/\SO(4)$, i.e., on the upper sheet of the
hyperboloid $x^2_0-x^2_1-x^2_2-x^2_3-x^2_4=1$, have the form
\cite{Var07}
\[
\fM^l_{mn}(\epsilon,\tau,\varepsilon,\omega)=e^{-m\epsilon}
\fP^l_{mn}(\cosh\tau)e^{-n(\varepsilon+\omega)}.
\]
Hence, in the two-particle case, we have the inner product
\[
\langle\boldsymbol{\Psi}_1\mid\boldsymbol{\Psi}_2\rangle=
\int\limits^\infty_0 d\tau d\epsilon d\varepsilon d\omega\sinh\tau
e^{-m\epsilon -n(\varepsilon +
\omega)}\fP^l_{m_1n_1}(\cosh\tau)\fP^l_{m_2n_2}(\cosh\tau).
\]
It is obvious that the integral
\[
I_1=\int\limits^\infty_0d\epsilon d\varepsilon d\omega e^{-m\epsilon
-n(\varepsilon + \omega)}
\]
converges. We calculate the integral
\begin{equation}\label{Inner1}
I_2=\int\limits^\infty_0\sinh\tau\fP^l_{m_1n_1}(\cosh\tau)
\fP^l_{m_2n_2}(\cosh\tau)d\tau.
\end{equation}
Using (\ref{Jacobi}), we express the functions
$\fP^l_{m_1n_1}(\cosh\tau)$ and $\fP^l_{m_2n_2}(\cosh\tau)$ in terms
of the hypergeometric functions. Then
\begin{multline}
I_2=\frac{1}{\Gamma(m_1-n_1+1)\Gamma(m_2-n_2+1)}\times\\
\sqrt{\frac{\Gamma(l-n_1+1)\Gamma(l+m_1+1)\Gamma(l-n_2+1)\Gamma(l+m_2+1)}
{\Gamma(l-m_1+1)\Gamma(l+n_1+1)\Gamma(l-m_2+1)\Gamma(l+n_2+1)}}\times
\nonumber
\end{multline}
\begin{multline}
\int\limits^\infty_0\cosh^{m_1+m_2+n_1+n_2}\frac{\tau}{2}
\sinh^{m_1+m_2-n_1-n_2}\frac{\tau}{2}\times\\
\hypergeom{2}{1}{l+m_1+1,m_1-l}{m_1-n_1+1}{-\sinh^2\frac{\tau}{2}}\times\\
\hypergeom{2}{1}{l+m_2+1,m_2-l}{m_2-n_2+1}{-\sinh^2\frac{\tau}{2}}\sinh\tau
d\tau.\label{Inner2}
\end{multline}
The first function $\fP^l_{m_1n_1}(\cosh\tau)$ can be written as
\begin{multline}
\fP^l_{m_1n_1}(\cosh\tau)=\sqrt{\frac{\Gamma(l-m_1+1)\Gamma(l-n_1+1)}
{\Gamma(l+m_1+1)\Gamma(l+n_1+1)}}
\cosh^{m_1+n_1}\frac{\tau}{2}\times\\
\sinh^{m_1-n_1}\frac{\tau}{2}\sum^{l-m_1}_{s=0}
\frac{(-1)^s\Gamma(l+m_1+s+1)\sinh^{2s}\frac{\tau}{2}}{\Gamma(s+1)
\Gamma(m_1-n_1+s+1)\Gamma(l-m_1-s+1)}.
\label{FpB}
\end{multline}
Taking (\ref{FpB}) into account, we rewrite the integral
(\ref{Inner2}) as
\begin{multline}
I_2=\frac{1}{\Gamma(m_2-n_2+1)}\times\\
\sqrt{\frac{\Gamma(l-m_1+1)\Gamma(l-n_1+1)\Gamma(l-n_2+1)\Gamma(l+m_2+1)}
{\Gamma(l+m_1+1)\Gamma(l+n_1+1)\Gamma(l-m_2+1)\Gamma(l+n_2+1)}}\times\\
\sum^{l-m_1}_{s=0}\frac{(-1)^s\Gamma(l+m_1+s+1)}{\Gamma(s+1)
\Gamma(m_1-n_1+s+1)\Gamma(l-m_1-s+1)}\times
\nonumber
\end{multline}
\begin{multline}
\int\limits^\infty_0\cosh^{m_1+m_2+n_1+n_2}\frac{\tau}{2}
\sinh^{m_1+m_2-n_1-n_2+2s}\frac{\tau}{2}\times\\
\hypergeom{2}{1}{l+m_2+1,m_2-l}{m_2-n_2+1}{-\sinh^2\frac{\tau}{2}}\sinh\tau
d\tau.\label{Inner3}
\end{multline}
Substituting $z=\cosh\tau$ in the integral
\begin{multline}
I_3=\int\limits^\infty_0\cosh^{m_1+m_2+n_1+n_2}\frac{\tau}{2}\times\\
\sinh^{m_1+m_2-n_1-n_2+2s}\frac{\tau}{2}
\hypergeom{2}{1}{l+m_2+1,m_2-l}{m_2-n_2+1}{-\sinh^2\frac{\tau}{2}}\sinh\tau
d\tau\nonumber
\end{multline}
we obtain
\begin{multline}
I_3=\int\limits^\infty_1\left(\frac{z^2-1}{4}\right)^{m_1+m_2}
\left(\frac{z+1}{z-1}\right)^{n_1+n_2}\times\\
\left(\frac{z-1}{2}\right)^s\hypergeom{2}{1}{l+m_2+1,m_2-l}{m_2-n_2+1}
{-\frac{z-1}{2}}dz.
\nonumber
\end{multline}
Further, introducing the new variable $t=-(z-1)/2$, we obtain
\begin{multline}
I_3=(-1)^{m_1+m_2+n_1+n_2+s+1}\times\\
\int\limits^\infty_0(1-t)^{m_1+m_2+n_1+n_2}
t^{m_1+m_2+s-n_1-n_2}\hypergeom{2}{1}{l+m_2+1,m_2-l}{m_2-n_2+1}{t}dt.
\nonumber
\end{multline}
Decomposing $(1-t)^{m_1+m_2+n_1+n_2}$ according to the Newton
binomial formula, we obtain
\begin{multline}
I_3=\sum^{m_1+m_2+n_1+n_2}_{p=0}(-1)^{m_1+m_2+n_1+n_2+s+p+1}\times\\
\frac{(m_1+m_2+n_1+n_2)!}{p!(m_1+m_2+n_1+n_2-p)!}\times\\
\int\limits^\infty_0t^{m_1+m_2+s+p-n_1-n_2}
\hypergeom{2}{1}{l+m_2+1,m_2-l}{m_2-n_2+1}{t}dt.\nonumber
\end{multline}
To calculate this integral, we use the formula \cite{Prud}:
\begin{multline}
I_4=\int t^n\hypergeom{2}{1}{a,b}{c}{t}dt=\\
=n!\sum^{n+1}_{k=1}(-1)^{k+1}\frac{(c-k)_kt^{n-k+1}}{(a-k+1)!(a-k)_k(b-k)_k}
\hypergeom{2}{1}{a-k,b-k}{c-k}{t}.\nonumber
\end{multline}
Then
\begin{multline}
I_3=\sum^{m_1+m_2+n_1+n_2}_{p=0}\sum^{m_1+m_2+s+p-n_1-n_2+1}_{k=1}
(-1)^{m_1+m_2+n_1+n_2
+s+p+k}\times\\
(m_1+m_2+s+p-n_1-n_2)!\times \nonumber
\end{multline}
\begin{multline}
\frac{(m_1+m_2+n_1+n_2)!(m_2-n_2-k+1)_k}
{(l+m_1-k+2)!(l+m_2-k+1)_k(m_2-l-k)_k}\times\\
t^{m_1+m_2+s+p-n_1-n_2-k+1}\hypergeom{2}{1}{l+m_2-k+1,m_2-l-k}{m_2-n_2-k+1}{t}.
\nonumber
\end{multline}
Taking (\ref{Inner3}) into account, we obtain the expression
\begin{multline}
\langle\boldsymbol{\Psi}_1\mid\boldsymbol{\Psi}_2\rangle=
\frac{1}{\Gamma(m_2-n_2+1)}\times\\
\sqrt{\frac{\Gamma(l-m_1+1)\Gamma(l-n_1+1)\Gamma(l-n_2+1)\Gamma(l+m_2+1)}
{\Gamma(l+m_1+1)\Gamma(l+n_1+1)\Gamma(l-m_2+1)\Gamma(l+n_2+1)}}\times
\nonumber
\end{multline}
\begin{multline}
\sum^{l-m_1}_{s=0}\sum^{m_1+m_2+n_1+n_2}_{p=0}
\sum^{m_1+m_2+s+p-n_1-n_2+1}_{k=1}(-1)^{m_1+m_2+n_1+n_2
+p+k}\times\\
\frac{\Gamma(l+m_1+s+1)(m_1+m_2+n_1+n_2)!}
{\Gamma(s+1)\Gamma(m_1-n_1+s+1)\Gamma(l-m_1-s+1)(l+m_2-k+2)!}\times
\nonumber
\end{multline}
\begin{multline}
\frac{(m_2-n_2-k+1)_k}{(l+m_2-k+1)_k(m_2-l-k)_k}\times\\
t^{m_1+m_2+s+p-n_1-n_2-k+1}\hypergeom{2}{1}{l+m_2-k+1,m_2-l-k}{m_2-n_2-k+1}{t}
\nonumber
\end{multline}
for the two-particle inner product. To investigate the convergence
of $\langle\boldsymbol{\Psi}_1\mid\boldsymbol{\Psi}_2\rangle$, we
apply the asymptotic expansion for the hypergeometric function
\cite{Bat}
\begin{multline}
\hypergeom{2}{1}{a,b}{c}{t}=\frac{\Gamma(c)\Gamma(b-a)}{\Gamma(b)
\Gamma(c-a)}(-t)^{-a}
\hypergeom{2}{1}{a,1-c+a}{1-b+a}{\frac{1}{t}}+\\
+\frac{\Gamma(c)\Gamma(a-b)}{\Gamma(a)\Gamma(c-b)}(-t)^{-b}
\hypergeom{2}{1}{b,1-c+b}{1-a+b}{\frac{1}{t}} .\nonumber
\end{multline}
Therefore,
\begin{multline}
\langle\boldsymbol{\Psi}_1\mid\boldsymbol{\Psi}_2\rangle=
\frac{1}{\Gamma(m_2-n_2+1)}\times\\
\sqrt{\frac{\Gamma(l-m_1+1)\Gamma(l-n_1+1)\Gamma(l-n_2+1)\Gamma(l+m_2+1)}
{\Gamma(l+m_1+1)\Gamma(l+n_1+1)\Gamma(l-m_2+1)\Gamma(l+n_2+1)}}\times
\nonumber
\end{multline}
\begin{multline}
\sum^{l-m_1}_{s=0}\sum^{m_1+m_2+n_1+n_2}_{p=0}
\sum^{m_1+m_2+s+p-n_1-n_2+1}_{k=1}(-1)^{m_1+m_2+n_1+n_2
+p+k}\times\\
\frac{\Gamma(l+m_1+s+1)(m_1+m_2+n_1+n_2)!}
{\Gamma(s+1)\Gamma(m_1-n_1+s+1)\Gamma(l-m_1-s+1)(l+m_2-k+2)!}\times\\
\frac{(m_2-n_2-k+1)_k}{(l+m_2-k+1)_k(m_2-l-k)_k}\times \nonumber
\end{multline}
\begin{multline}
\left[(-1)^{l+m-k+1}\frac{\Gamma(m_2-n_2-k+1)\Gamma(-2l-1)}{\Gamma(m_2-l-k)
\Gamma(-n_2-l)}\right.\times\\
t^{m_1+s+p-n_1-n_2-l}\hypergeom{2}{1}{l+m_2-k+1,l+n_2+1}{2l+2}{\frac{1}{t}}+
\nonumber
\end{multline}
\begin{multline}
(-1)^{m_2-l-k}\frac{\Gamma(m_2-n_2-k+1)\Gamma(2l+1)}{\Gamma(l+m_2-k+1)
\Gamma(-n_2-l)}\times\\
\left.t^{m_1+l+s+p-n_1-n_2+1}\hypergeom{2}{1}{m_2-l-k,n_2-l}{2-2l}
{\frac{1}{t}}\right].
\nonumber
\end{multline}
In this expression the hypergeometric function ${}_2F_1$ can be
written as a power series in $1/t$. It hence follows that
$\langle\boldsymbol{\Psi}_1\mid\boldsymbol{\Psi}_2\rangle\sim
t^{m_1+l+s+p-M-n_1-n_2+1}$, and because $M\rightarrow\infty$,
$\langle\boldsymbol{\Psi}_1\mid\boldsymbol{\Psi}_2\rangle$ converges
for $M>m_1+l+s+p-n_1-n_2+1$.

\section{Summary}
We have presented an extended group-averaging method by determining
the integrals giving the inner products of free fields on
homogeneous spaces of the de Sitter group $\SO_0(1,4)$. We
considered $N$-particle case on the four-dimensional hyperboloid
$H^4$. It would be interesting to consider inner products and also
their convergence on other homogeneous spaces of $\SO_0(1,4)$ (both
symmetric Riemannian and non-Riemannian), such as the
three-dimensional real sphere $S^3$, the two-dimensional quaternion
sphere $S^q_2$, and the group manifold $\fS_{10}$ of $\SO_0(1,4)$.
Our next paper will be devoted to these questions.

\end{document}